\documentclass[12pt]{article}
\usepackage{amssymb,amsfonts}
\usepackage{graphics,psboxit,amsmath}
\usepackage{subfigure}
\usepackage{graphicx}
\usepackage{verbatim}
\usepackage{color}

\definecolor{markcolor}{rgb}{.25,0,1}

\definecolor{markcolor2}{rgb}{1,0,0}

\definecolor{markcolor3}{rgb}{0,1,0}

\def\hybrid{\topmargin -30pt    \oddsidemargin 0pt 
        \headheight 0pt \headsep 0pt
        \textwidth 6.25in       
        \textheight 9.5in       
        \marginparwidth .875in
        \parskip 5pt plus 1pt   \jot = 1.5ex}

\hybrid

\catcode`\@=11

\def\marginnote#1{}
\newcount\hour
\newcount\minute
\newtoks\amorpm
\hour=\time\divide\hour by60 \minute=\time{\multiply\hour by60
\global\advance\minute by-\hour}
\edef\standardtime{{\ifnum\hour<12 \global\amorpm={am}%
        \else\global\amorpm={pm}\advance\hour by-12 \fi
        \ifnum\hour=0 \hour=12 \fi
        \number\hour:\ifnum\minute<10 0\fi\number\minute\the\amorpm}}
\edef\militarytime{\number\hour:\ifnum\minute<10 0\fi\number\minute}

\def\draftlabel#1{{\@bsphack\if@filesw {\let\thepage\relax
   \xdef\@gtempa{\write\@auxout{\string
      \newlabel{#1}{{\@currentlabel}{\thepage}}}}}\@gtempa
   \if@nobreak \ifvmode\nobreak\fi\fi\fi\@esphack}
        \gdef\@eqnlabel{#1}}
\def\@eqnlabel{}
\def\@vacuum{}
\def\draftmarginnote#1{\marginpar{\raggedright\scriptsize\tt#1}}

\def\draft{\oddsidemargin -.5truein
        \def\@oddfoot{\sl preliminary draft \hfil
        \rm\thepage\hfil\sl\today\quad\militarytime}
        \let\@evenfoot\@oddfoot \overfullrule 3pt
        \let\label=\draftlabel
        \let\marginnote=\draftmarginnote
   \def\@eqnnum{(\theequation)\rlap{\kern\marginparsep\tt\@eqnlabel}%
\global\let\@eqnlabel\@vacuum}  }

\def\draft2{
        \def\@oddfoot{\sl preliminary draft \hfil
        \rm\thepage\hfil\sl\today\quad\militarytime}
        \let\@evenfoot\@oddfoot \overfullrule 3pt
        \let\label=\draftlabel
        \let\marginnote=\draftmarginnote
   \def\@eqnnum{(\theequation)\rlap{\kern\marginparsep\tt\@eqnlabel}%
\global\let\@eqnlabel\@vacuum}  }

\def\preprint{\twocolumn\sloppy\flushbottom\parindent 2em
        \leftmargini 2em\leftmarginv .5em\leftmarginvi .5em
        \oddsidemargin -.5in    \evensidemargin -.5in
        \columnsep .4in \footheight 0pt
        \textwidth 10.in        \topmargin  -.4in
        \headheight 12pt \topskip .4in
        \textheight 6.9in \footskip 0pt
        \def\@oddhead{\thepage\hfil\addtocounter{page}{1}\thepage}
        \let\@evenhead\@oddhead \def\@oddfoot{} \def\@evenfoot{} }

\def\numberbysection{\@addtoreset{equation}{section}
        \def\theequation{\thesection.\arabic{equation}}}

\def\underline#1{\relax\ifmmode\@@underline#1\else
        $\@@underline{\hbox{#1}}$\relax\fi}

\def\titlepage{\@restonecolfalse\if@twocolumn\@restonecoltrue\onecolumn
     \else \newpage \fi \thispagestyle{empty}\c@page\z@
        \def\thefootnote{\fnsymbol{footnote}} }

\def\endtitlepage{\if@restonecol\twocolumn \else \newpage \fi
        \def\thefootnote{\arabic{footnote}}
        \setcounter{footnote}{0}}  

\catcode`@=12 \relax

\def\figcap{\section*{Figure Captions\markboth
        {FIGURECAPTIONS}{FIGURECAPTIONS}}\list
        {Figure \arabic{enumi}:\hfill}{\settowidth\labelwidth{Figure
999:}
        \leftmargin\labelwidth
        \advance\leftmargin\labelsep\usecounter{enumi}}}
 \relax
\def\tablecap{\section*{Table Captions\markboth
        {TABLECAPTIONS}{TABLECAPTIONS}}\list
        {Table \arabic{enumi}:\hfill}{\settowidth\labelwidth{Table
999:}
        \leftmargin\labelwidth
        \advance\leftmargin\labelsep\usecounter{enumi}}}
 \relax
\def\reflist{\section*{References\markboth
        {REFLIST}{REFLIST}}\list
        {[\arabic{enumi}]\hfill}{\settowidth\labelwidth{[999]}
        \leftmargin\labelwidth
        \advance\leftmargin\labelsep\usecounter{enumi}}}
 \relax
\makeatletter
\newcounter{pubctr}
\def\publist{\@ifnextchar[{\@publist}{\@@publist}}
\def\@publist[#1]{\list
        {[\arabic{pubctr}]\hfill}{\settowidth\labelwidth{[999]}
        \leftmargin\labelwidth
        \advance\leftmargin\labelsep
        \@nmbrlisttrue\def\@listctr{pubctr}
        \setcounter{pubctr}{#1}\addtocounter{pubctr}{-1}}}
\def\@@publist{\list
        {[\arabic{pubctr}]\hfill}{\settowidth\labelwidth{[999]}
        \leftmargin\labelwidth
        \advance\leftmargin\labelsep
        \@nmbrlisttrue\def\@listctr{pubctr}}}
 \relax
\makeatother

\def\be{\begin{equation}}
\def\ee{\end{equation}}
\def\ba{\begin{eqnarray}}
\def\ea{\end{eqnarray}}

\def\del{\partial}

\newcommand{\fr}[1]{\mathfrak{#1}}

\def\d{\delta}

\def\m{\mu}

\def\l{\lambda}

\def\s{\sigma}

\def\TT{{\mathcal{T}}}

\def\no{\noindent}

\def\qq{\qquad}

\def\IR{\relax{\rm I\kern-.18em R}}

\def\bse{\begin{small}\begin{equation*}}
\def\ese{\end{equation*}\end{small}}

\begin{document}


\renewcommand{\theequation}{\thesection.\arabic{equation}}
\csname @addtoreset\endcsname{equation}{section}

\newcommand{\eqn}[1]{(\ref{#1})}

\begin{titlepage}\strut\hfill
\vskip 1.3cm
\begin{center}


{\large \bf Generalized Landau-Lifshitz models on the interval}

\vskip 0.5in

{\bf Anastasia Doikou} and {\bf Nikos Karaiskos} \vskip 0.1in

 {\footnotesize Department of Engineering Sciences, University of Patras, Physics Division\\
GR-26500 Patras, Greece}

\vskip .1in


{\footnotesize {\tt E-mail:$\{$adoikou, nkaraiskos$\}@$upatras.gr}}\\

\end{center}

\vskip .4in

\centerline{\bf Abstract}

We study the classical generalized $\mathfrak{gl}_n$ Landau-Lifshitz (L-L) model with special boundary conditions that preserve integrability.
We explicitly derive the first non-trivial local integral of motion, which corresponds to the boundary
Hamiltonian for the $\mathfrak{sl}_2$ L-L model. Novel expressions of the modified Lax pairs associated to the integrals of
motion are also extracted. The relevant equations of motion with the corresponding boundary conditions are
determined. Dynamical integrable boundary conditions are also examined within this spirit. Then the generalized isotropic and anisotropic $\mathfrak{gl}_n$ Landau-Lifshitz models are considered, and novel expressions of the boundary Hamiltonians and the relevant equations of motion and boundary conditions are derived.
\no

\vfill

\end{titlepage}
\vfill \eject


\tableofcontents

\section{Introduction}

Numerous investigations have been devoted to the issue of incorporating non-trivial conditions that preserve integrability both in discrete \cite{Sklyanin:1987bi}--\cite{gama}, and continuum integrable systems \cite{Sklyanin:1987bi},  \cite{cherednik}--\cite{doikou1}.
The central purpose of the present article is the study of classical
integrable models when general boundaries that preserve integrability are implemented. Among the various
classes of integrable models we choose to consider here a prototype model, that is the generalized classical continuum Heisenberg or Landau-Lifshitz (L-L) model. This model may be thought of as the immediate classical analogue of the XXX (XYZ for the anisotropic case) quantum spin chain \cite{FR, ADS}, whereas higher rank generalizations may be seen as continuum limits of known high rank quantum spin chains. Although much attention has been devoted to the investigation of the quantum models with integrable boundary conditions not much progress has been made --from the algebraic point of view-- on their classical continuum counterparts. Here, we consider the classical continuum case, we identify the boundary Hamiltonian, and the relevant boundary Lax pair, for the $\mathfrak{sl}_2$ L-L model, utilizing primarily the algebra that rules the model, that is the classical reflection algebra. Our study concerns not only typical $c$-number reflection matrices, but also dynamical reflection matrices, which give rise to dynamical type boundary conditions for the aforementioned model. The generalized $\mathfrak{gl}_n$ L-L models are also examined within this spirit.

It is worth noting that the significance of the particular study stems primarily from the fact that it provides novel results for a wide class of classical integrable models associated to the $\mathfrak{gl}_n$ algebra. Note that such anisotropic (trigonometric) models may be appropriately mapped to $A_{n-1}^{(1)}$ affine Toda field theories (see e.g. \cite{Faddeev:1987ph}). Moreover, the present investigation provides a first systematic description on the issue of integrable continuum limits of discrete integrable models, that contain boundary type terms or other distinct local terms, such as the ones arising also in the case of integrable defects.

The outline of the article is as follows: in the next section we briefly review the model with periodic boundary conditions, as well as the relevant fundamental ingredients (see also \cite{Faddeev:1987ph}). The Lax pair formulation, and the construction of the associated integrals of motion through the associated algebras are reviewed. In 
section 3 we review Sklyanin's generic algebraic frame \cite{Sklyanin:1987bi} describing classical models with boundaries that preserve integrability. Based on this framework we explicitly derive the associated Hamiltonian with suitable integrable boundary terms recovering also some of the expressions presented in \cite{Sklyanin:1987bi}. Note that in \cite{Sklyanin:1987bi} only diagonal boundary terms were treated, whereas here the most general boundary terms that preserve integrability are derived. Dynamical boundaries are also examined within this context. Note that our results are consistent with the classical continuum limits of the relevant quantum discrete Hamiltonians. In section 4 we review the construction of modified Lax pairs in the presence of integrable boundaries discussed in \cite{Avan:2007sf}. Relying on this framework, and using generic solutions of the reflection equation --$c$-number and dynamical-- we are able to derive the Lax pair associated to the extracted Hamiltonians. This way the consistency of the whole procedure is fully ensured. In section 5
the continuum limit of the XXZ open spin chain is considered leading to the boundary antitropic $\mathfrak{sl}_2$ L-L model. This further ensures the validity of the continuum limit process followed.
We then examine the isotropic and anisotropic $\mathfrak{gl}_n$ L-L models as continuum limits of the $\mathfrak{gl}_n$ and  ${\mathfrak U}_q(\mathfrak{gl}_n)$ open spin chains respectively, and obtain the classical continuum Hamiltonians, and the associated equations of motion and boundary conditions.

\section{The isotropic $\mathfrak{sl}_2$ Landau-Lifshitz model}
Let us briefly review the continuous isotropic Landau-Lifshitz (L-L) model with periodic boundary conditions, associated to the $\mathfrak{sl}_2$ classical algebra (see also \cite{Faddeev:1987ph}).
After introducing the basic ingredients of the model and setting up our notations, we recall the Lax pair formulation
for the classical integrable Hamiltonian system, and discuss the systematic means of
constructing the whole tower of local integrals in involution.

The physical quantities of the model are described by vector-valued functions $\vec{S}(x)=
(S_1(x),S_2(x),S_3(x))$ taking values on the unit 2-sphere
\be
\vec{S}^2(x) = \sum_{i=1}^3 S_i^2(x) = 1.
\ee
Note that throughout the text we shall also use the following combinations of $S_i(x)$
\be
S^{\pm}(x) = \frac{1}{2}(S_1 (x) \pm i S_2(x)).
\ee
The equations of motion associated to the isotropic Landau-Lifshitz model, 
which is our main interest here, are of the form:
\be
\frac{\del \vec{S}}{\del t}=i \vec{S} \wedge {\partial^2 \vec{S} \over \partial x^2}.
\label{e.o.m.}
\ee
The fields $S_i(x)$ obey boundary conditions, which are taken to be such that either $S_i(x)$
become periodic, i.e. $S_i(x+2L) = S_i(x)$, or consider the fields and their derivatives to be zero at the endpoints (Schwartz boundary conditions).

The Poisson structure of the phase space for the physical quantities $S_i(x)$ is given by the
Poisson brackets
\be
\{ S_a(x),S_b(y)\} = 2i\varepsilon_{abc}S_c(x)\d(x-y),
\ee
where $\varepsilon_{abc}$ is the totally antisymmetric Levi-Civita tensor with value $\varepsilon_{123} = 1$.
The Hamiltonian of the model is given by
\be
H=- \frac{1}{4}\int\left(\left(\frac{\del S_1}{\del x}\right)^2 + \left(\frac{\del S_2}{\del x}\right)^2
+ \left(\frac{\del S_3}{\del x}\right)^2 \right)dx.
\ee
The equations of motion in the Hamiltonian form are expressed as
\be
\frac{\del \vec{S}}{\del t} = \{H,~\vec{S}\}. \label{emh}
\ee
Other physical integrals of motion include the momentum, which is given by
\be
P=\int \frac{S_1\frac{\del S_2}{\del x}-S_2 \frac{\del S_1}{\del x}}{1+S_3}dx,
\ee
and the total spin of the model in the case where periodic boundaries are considered.

\subsection{The Lax pair formulation}
Within the Lax pair formulation of a classical integrable Hamiltonian system one first defines the auxiliary
linear differential problem, which reads as
\ba
\frac{\del}{\del x}\Psi(x,t) & = & \mathbb{U}(x,t,\l)\Psi(x,t) \cr \label{eq1}
\frac{\del}{\del t}\Psi(x,t) & = & \mathbb{V}(x,t,\l)\Psi(x,t).
\ea
In general, $\mathbb{U}$ and $\mathbb{V}$ are $n\times n$ matrices. Their entries contain dynamical fields, their
derivatives and possibly, the spectral parameter $\l$. The compatibility condition of these two equations
leads to the so-called zero curvature condition
\be
\del_t\mathbb{U}-\del_x\mathbb{V}+[\mathbb{U},~\mathbb{V}]=0,
\label{zero_curv}
\ee
which provides the equations of motion of the system under consideration.

One then constructs the monodromy matrix
\be
T(x,y,\l)=\mathrm{P}exp\left(\int_y^x\mathbb{U}(z)dz\right),
\label{monodromy}
\ee being a solution of the equation (\ref{eq1}).
Assume that
$\mathbb{U}$ obeys the classical linear Poisson algebraic relation \cite{Faddeev:1987ph}
\be
\{\mathbb{U}_a(x,\l),~\mathbb{U}_b(y,\m)\} = [r_{ab}(\l-\m),~\mathbb{U}_a(x,\l)+\mathbb{U}_b(y,\m)]
\d(x-y),
\label{pois_cl_int}
\ee
then it follows that the monodromy matrix satisfies the quadratic algebraic relation,
\be
\{T_a(x,y,t,\l),~T_b(x,y,t,\m)\}=[r_{ab}(\l-\m),~T_a(x,y,t,\l)T_b(x,y,t,\m)].
\label{monod_pois}
\ee
$r_{ab}$ is the classical $r$-matrix corresponding to the Hamiltonian system, and
satisfies the classical Yang-Baxter equation \cite{sts}. The conserved charges may be
obtained via the expansion of $t(\l)=tr T(\l)$ in powers of the spectral parameter, $\l$. It can also
be shown via (\ref{monod_pois}) that these charges are in involution, that is they satisfy \cite{Faddeev:1987ph}
\be
\{t(\l),~t(\m)\} = 0.
\ee

In the case of the L-L model, the classical $r$-matrix has the simple form \cite{yang}
\be
r(\l) = \frac{\mathcal{P}}{\l},
\ee
where $\mathcal{P}$ is the permutation operator: $\mathcal{P}(\vec{a}\otimes \vec{b}) = \vec{b}\otimes \vec{a}$.

We shall restrict ourselves for the moment in the case of the $\fr{sl}_2$ L-L model. In this case,
the permutation and Lax operators are respectively
\be
\mathcal{P} =
\begin{pmatrix}
 1 & 0 & 0 & 0 \cr
 0 & 0 & 1 & 0 \cr
 0 & 1 & 0 & 0 \cr
 0 & 0 & 0 & 1 \cr
\end{pmatrix},
\qquad
\mathbb{U}(x) =  \frac{1}{\l}
\begin{pmatrix}
 \frac{S_3}{2} & S^- \cr
 S^+ & -\frac{S_3}{2}
\end{pmatrix}
\equiv \frac{1}{2\l}\mathcal{S}.
\label{U_bulk}
\ee
We also note here the explicit form of the $\mathbb{V}$-operator
\be
\mathbb{V}(x) =  \frac{1}{2  \l^2} \mathcal{S} - \frac{1}{2\l} \frac{\del\mathcal{S}}{\del x}
\mathcal{S}.
\label{V_bulk}
\ee
Inserting the Lax pair operators \eqn{U_bulk} and \eqn{V_bulk} into the zero curvature condition
\eqn{zero_curv} yields exactly the equations of motion \eqn{e.o.m.}.

\subsection{Integrals of motion}
As already mentioned, the local integrals of motion may be extracted through the expansion of
the trace of the monodromy matrix in powers of the spectral parameter $\l$. A key property of the
monodromy matrix that is crucial in what follows is that is satisfies:
\be
\frac{\del}{\del x} T(x,y,\l)= \mathbb{U}(x,\l)T(x,y,\l).
\label{tr_eq}
\ee
Let us consider the following ansatz for the monodromy matrix:
\be
T(x,y,\l) = (1+W(x,\l))e^{Z(x,y,\l)}(1+W(y,\l))^{-1},
\label{transatz}
\ee
$W$ and $Z$ are purely off-diagonal and diagonal matrices respectively. We also assume that $W,\ Z$
are expressed as:
\be
W(x,\l)=\sum_{n=0}^{\infty}\l^n W_n(x), \qq Z(x,y,\l)=\sum_{n=-1}^{\infty}\l^n Z_n(x,y).
\label{WZexp}
\ee
Our main aim henceforth is to identify the elements $W_n,\ Z_n$, and hence the integrals of motion.
It is technically convenient to split the Lax operator into a diagonal and an off-diagonal part as
\be
\mathbb{U} = \mathbb{U}_d + \mathbb{U}_a \equiv \frac{1}{2\l}
\begin{pmatrix}
  S_z & 0 \cr 0 & -S_z
\end{pmatrix}
+ \frac{1}{\l}
\begin{pmatrix}
  0 & S^- \cr S^+ & 0
\end{pmatrix}.
\ee
Substituting the ansatz \eqn{transatz} into the relation \eqn{tr_eq}, and splitting the
resulting equation into a diagonal and an off-diagonal part one obtains
\ba
&& \frac{dW}{dx} + W \mathbb{U}_d - \mathbb{U}_d W +  W \mathbb{U}_a W- \mathbb{U}_a =0,\cr
&& \frac{\del Z}{\del x} = \mathbb{U}_d + \mathbb{U}_a W.
\ea
Plugging in the explicit expressions
for $\mathbb{U}_a,\mathbb{U}_d$, we end up with the following relation for $W$
\be
\frac{dW}{dx} + \frac{1}{\l}W S_3~\s^3 - \frac{1}{\l}(S^-\s^+ + S^+\s^-) +
\frac{1}{\l}W(S^-\s^+ + S^+\s^-)W = 0,
\ee
with $\s^3,\ \sigma^{\pm}$ being the familiar $2\times 2$ Pauli matrices:
\be
\sigma^3=\begin{pmatrix}1 & 0 \\ 0 & -1\end{pmatrix}, \qquad \sigma
^+=\begin{pmatrix} 0 & 1 \\ 0 & 0 \end{pmatrix},
\qquad \sigma^-=\begin{pmatrix}0 & 0 \\ 1 & 0 \end{pmatrix}.
\label{pauli}
\ee
Inserting the expansion \eqn{WZexp} in the equation above, one determines all $W_n$'s. We
are only interested in the first three terms of the expansion, which are given by:
\ba
&& \mathcal{O}(1/\l): \qquad
W_0 = \begin{pmatrix}
       0 & -\bar{a} \cr a & 0
      \end{pmatrix}, \qquad a = \frac{1-S_3}{2S^-} = \frac{2 S^+}{1+S_3},\cr
&& \mathcal{O}(\l^{0}): \qquad
W_1 = \begin{pmatrix}
       0 & -\bar{a}' \cr -a' & 0
      \end{pmatrix},\cr
&& \mathcal{O}(\l): \qquad
W_2 = \begin{pmatrix}
       0 & -\bar{a}''+(\bar{a}')^2S^+ \cr a''-(a')^2S^- & 0
      \end{pmatrix}
\equiv \begin{pmatrix}
       0 & -\bar{b} \cr b & 0
      \end{pmatrix}.
\ea
The first
three terms suffice in order to compute the first two integrals of motion, namely
the momentum and the Hamiltonian. To complete the computation we should also determine $Z$, provided by the
following equation
\be
\frac{\del Z}{\del x} = \mathbb{U}_d + \mathbb{U}_a W.
\ee
Substituting the relevant expansion of $W$ and the explicit forms of $\mathbb{U}_d,
\mathbb{U}_a$ into the equation above we conclude
\be
Z(x,y,\l)=\frac{1}{2\l}(x-y)\s_3 + \sum_{n=1}^{\infty}\l^{n-1}
\int_y^x(S^+\s^-+S^-\s^+)W_n(z) dz.
\ee

By using the exact expressions for $W_n$, one may finally determine $Z$, order by order. In particular,
as noted above, at order $\mathcal{O}(1/\l)$ we found
\be
Z_{-1}=\frac{1}{2}(x-y)\s^3.
\ee
Here we set $x= L$ and $y =-L$.
The explicit form of $Z_{-1}$ is important since it indicates the leading contribution of $e^Z$ as
$\l\to0$, a result which will be greatly used below and in the forthcoming sections. Moving on to the
next orders, one naturally arrives at the first two integrals of motion, as expected. More specifically,
by working with the conventions defined above one concludes that
\be
(Z_0)_{11}\propto P, \qquad \textrm{and} \qquad (Z_1)_{11} \propto H.
\ee
Extra care is needed while computing the integrals of
motion above, taking into account the Schwartz boundary conditions at the endpoints for the
fields and their derivatives, namely $(S_i(\pm L),S'_i(\pm L))\to 0$.
Note that in the case of non-trivial boundaries extra terms, emerging from the bulk part, should be taken into account,
given that the fields do not vanish in that case at the endpoints. This will be transparent in the subsequent sections.

\section{Implementing integrable boundaries}
We shall now briefly describe the relevant machinery needed to implement special
boundaries into a classical model in a way that integrability is ensured.
After reviewing the general setting we focus on the particular example of interest,
that is the Landau-Lifshitz model with integrable boundaries.

\subsection{Algebraic setting}
We review Sklyanin's formulation \cite{Sklyanin:1987bi} in order to exhibit
the integrability for classical
models with special boundaries. Suppose that we have the monodromy matrix of the model at hand, endowed
with the Poisson structure \eqn{monod_pois}. Let also $K^{\pm}(\l)$ be $c$-number (non-dynamical) representations of the classical reflection algebra \cite{Sklyanin:1987bi, cherednik},
\ba
&& 0 = [r_{12}(\l_1-\l_2),~ K(\l_1)\ K_2(\l_2)] \cr
&& +K_1(\l_1)\ r_{12}(\l_1+\l_2)\ K_2(\l_2) - K_2(\l_2)\ r_{12}(\l_1+\l_2)\ K_1(\l_1). \label{cnumber}
\ea
Our notation is such that $K^+(\l) =K(-\l,~\xi^+,~{\mathrm k}^-)$ and $K^-(\l)=K(\l,~\xi^-, ~{\mathrm k}^+)$; $~\xi^{\pm},\ {\mathrm k}^{\pm}$ are generic free boundary parameters (see also below in the text).
One may then define a modified
transition matrix as \cite{Sklyanin:1987bi}
\be
\TT(x,y,\l) = T(x,y,\l)~K^-(\l)~\hat{T}(x,y,\l),
\ee
where $\hat{T}(\l)=T^{-1}(-\l)$.
The modified monodromy matrix satisfies the classical version of the reflection equation \cite{Sklyanin:1987bi, cherednik}:
\ba
&& \left\{\TT_1(\l_1),~ \TT_2(\l_2)\right\} = [r_{12}(\l_1-\l_2),~ \TT(\l_1)\ \TT_2(\l_2)] \cr
&& +\TT_1(\l_1)\ r_{12}(\l_1+\l_2)\ \TT_2(\l_2) - \TT_2(\l_2)\ r_{12}(\l_1+\l_2)\ \TT_1(\l_1). \label{clrefl}
\ea
The generalized transfer matrix reads as
\be
t(x,y,\l)=tr\{K^+(\l)~\TT(x,y,\l)\}, \label{opentransfer}
\ee
and it is shown to satisfy \cite{Sklyanin:1987bi}
\be
\left\{ t(x,y,\l_1),~ t(x,y,\l_2) \right\} =0, \qquad \l_1,\l_2 \in \mathbb{C},
\ee
hence it may be interpreted as the generating functional of the conserved integrals of motion.

By adopting the ansatz \eqn{transatz} and setting $x=0$, $y=-L$ for the boundary
points, the generating functional of the {\it local} integrals of motion takes the form
\[
\ln~tr\Big \{K^+(\l)\ T(0,-L,\l)\ K^-(\l)\ \hat{T}(0,-L,\l)\Big \} =
\]
\be
\ln ~tr\{(1+\hat{W}(0))^{-1}K^+(\l)(1+W(0))\ e^{Z(0,-L)} (1+W(-L))^{-1}K^-(\l) (1+\hat{W}(-L))
e^{-\hat{Z}(0,-L)}\},
\label{mod_trans}
\ee
where $\hat W$ are the same as before, but with $\l \to -\l$. It is this expression
that one expands in powers of $\l$, in order to derive the local integrals of motion.

\subsection{The boundary Hamiltonian}
We now proceed in deriving the integrals of motion for the Landau-Lifshitz model with
integrable boundaries. We shall be using the following $c$-number representation of the classical reflection algebra \cite{deve}
\be
K(\l,~\xi,~{\mathrm k})=
\begin{pmatrix}
 -\l +i\xi & 2{\mathrm k}\l \cr 2{\mathrm k}\l & \l +i\xi
\end{pmatrix}.
\ee
As in the periodic  case to derive the integrals of motion, one expands the generic object \eqn{mod_trans} in powers
of $\l$. We shall only present here the final results, whereas the technical details of the derivation can
be found in the Appendix A. One should keep in mind that extra terms, emerging directly from the bulk,
cancel out suitably some purely boundary terms providing eventually a quite simple boundary contribution to the Hamiltonian.

In the case of open boundary conditions, the first integral of motion becomes
trivial. This is actually expected, given that this is essentially the momentum, which is not a
conserved quantity anymore. The second integral of motion, i.e. the Hamiltonian is
computed to be (see also \cite{Sklyanin:1987bi}, where only diagonal boundary terms are considered)
\ba
\mathcal{I}_1 = -\frac{1}{4}\int_{-L}^0\left(\frac{\del \vec{S}}{\del x}\right)^2dx
-\frac{i}{2\xi^-}\Big (2{\mathrm k}^-S_1(-L) - S_3(-L)\Big )
+ \frac{i}{2\xi^+}\Big (2{\mathrm k}^+S_1(0) - S_3(0)\Big ). \label{bll}
\ea
The first term in the expression above is just the bulk term, while the rest are total boundary
contributions.

\subsubsection{Dynamical boundary conditions}
We shall now discuss the case of dynamical degrees of freedom, attached at the ends of the system. To achieve this we shall consider the dynamical solution of the reflection equation \cite{Sklyanin:1987bi}:
\be
{\mathbb K}(\lambda) = {\mathbb L}(\lambda)\ K(\lambda)\ {\mathbb L}^{-1}(-\lambda), \label{kdyn}
\ee
where we define
\be
{\mathbb L}= \lambda {\mathbb I}+
\begin{pmatrix}  \frac{{\mathbb S}_3}{2} & {\mathbb S}^- \cr
 {\mathbb S}^+ & -\frac{{\mathbb S}_3}{2}
\end{pmatrix}. \label{ll1}
\ee
The ${\mathbb L}$ matrix satisfies the quadratic algebra (\ref{monod_pois}) with the Yangian $r$-matrix.
The elements ${\mathbb S}^z,\ {\mathbb S}^{\pm}$ apparently satisfy the classical $\fr{sl}_2$ algebra. Note that special limits of the generic ${\mathbb L}$ matrix lead to the Discrete-Self-Trapping (DST) model, or the Toda model (see e.g. \cite{Doikou:2007mm} and references therein), so this way one treats a generic class of dynamical boundaries.

We shall consider here for simplicity, but without loss of generality, the dynamical boundary attached to the left end of the system. The right end of the system will be described by the trivial reflection matrix $K^+ \propto {\mathbb I}$. Hence, the generalized transfer matrix will be of the form:
\be
t(\lambda) =tr_a\Big [T_a(\lambda)\ {\mathbb K}^-_a(\lambda)\ T_a^{-1}(-\lambda) \Big ].
\ee
It will be convenient for the following computations to express the ${\mathbb K}$-matrix as:
\be
{\mathbb K}^-(\lambda) \propto {\mathbb I} + \lambda {\mathbb B} + {\cal O}(\lambda^2),
~~~~~~{\mathbb B} = \begin{pmatrix}-{\mathbb X} & {\mathbb Z} \cr
 {\mathbb Y} & {\mathbb X}
\end{pmatrix}, \label{dynamic}
\ee
where we define
\ba
&&{\mathbb X} =  -4 {\mathbb S}_3 + {1 \over i\xi^-} \Big ( 2{\mathbb S}_3^2 -1 -4 {\mathrm k}^- {\mathbb S}_3 ({\mathbb S}^+ +{\mathbb S}^-) \Big ) \cr
&& {\mathbb Y} = 8{\mathbb S}^+ +{1 \over i\xi^-} \Big ( 8{\mathrm k}^- ({\mathbb S}^{+})^2 -2 {\mathrm k}^- {\mathbb S}_3^2 -4 {\mathbb S_3} {\mathbb S}^+\Big ) \cr
&& {\mathbb Z} = 8{\mathbb S}^- + {1 \over i\xi}^- \Big ( 8{\mathrm k}^- ({\mathbb S}^{-})^2 -2 {\mathrm k}^- {\mathbb S}_3^2 -4 {\mathbb S_3} {\mathbb S}^-\Big ). \label{opers}
\ea
It is clear that the elements ${\mathbb X},\ {\mathbb Y},\ {\mathbb Z}$ contain the dynamical degrees of freedom attached to the boundary.
Expanding appropriately the generalized transfer matrix as in the previous section, we end up to the following Hamiltonian with distinct dynamical terms attached to one boundary:
\be
\mathcal{I}_1 = -\frac{1}{4}\int_{-L}^0\left(\frac{\del \vec{S}}{\del x}\right)^2dx
+\frac{1}{2}S^+(-L){\mathbb Z} +\frac{1}{2} S^-(-L){\mathbb Y} - \frac{1}{2}S_3(-L) {\mathbb X}. \label{hdyn}
\ee
Note that such boundary terms would have emerged for the other end of the system as well,
after implementing a similar right dynamical reflection matrix. More precisely, a dynamical $K^+$ matrix for the right boundary would lead to extra boundary terms in expressions (\ref{hdyn}) at $x=0$
of exactly the same from of as the ones at $x=-L$, but with $\xi^- \to -\xi^+,\ {\mathrm k}^- \to {\mathrm k}^+$ in (\ref{opers}).

\section{The modified Lax pair}
When integrable boundary conditions are implemented, the Lax pairs
associated to the integrals of motion are accordingly modified. The systematic construction of the modified Lax pairs was presented in \cite{Avan:2007sf}. In what follows, we briefly review the
results of \cite{Avan:2007sf}, and then apply the formalism in the case of the $\mathfrak{sl}_2$
Landau-Lifshitz model with integrable boundaries.

\subsection{Reviewing the construction}
Recall first the construction of the $\mathbb{V}$-operator associated to a given integral of motion
for a classical integrable model with periodic boundary conditions. Using \eqn{monod_pois} one formulates the following Poisson structure:
\be
\Big \{ T_a(L,
-L, \lambda),\  {\mathbb U}_b(x, \mu) \Big \}= {\partial M(x,
\lambda, \mu) \over \partial x} + \Big [ M(x, L, -L, \lambda,
\mu),\ {\mathbb U}_{b}(x, \mu) \Big ],
\label{M_pois}
\ee
where we define
\be
M(x, \lambda, \mu) = T_{a}(L, x, \lambda) r_{ab}(\lambda -\mu)\ T_a(x, -L, \lambda).
\label{M_def}
\ee
More details on the derivation of the latter formula can be found in \cite{Faddeev:1987ph}. Recalling now that $t(\l) =
tr T(\l)$ it naturally follows from \eqn{M_pois} and \eqn{zero_curv} that
\be
\Big \{ \ln\ t(\lambda),\ {\mathbb U}(x,
\lambda) \Big \} = {\partial {\mathbb V}(x, \lambda, \mu) \over
\partial x} + \Big [ {\mathbb V}(x, \lambda, \mu),\ {\mathbb
U}(x,\lambda) \Big ],
\ee
with
\be {\mathbb V}(x, \lambda, \mu) =
t^{-1}(\lambda)\ tr_a \Big ( T_a(L, x, \lambda)\ r_{ab}(\lambda,
\mu)\ T_a( x, -L, \lambda) \Big ).
\label{vv}
\ee

In the case of open boundary conditions one may prove that a generalized Poisson structure holds \cite{Avan:2007sf}, i.e.
\be
\Big \{{\cal
T}_a(0, -L,\lambda),\ {\mathbb U}_b(x, \mu) \Big \} = {\mathbb M}_a'(x,
\lambda, \mu) + \Big [{\mathbb M}_a(x, \lambda, \mu),\ {\mathbb
U}_b(x,\ \mu ) \Big],
\label{last}
\ee
where we now define
\ba {\mathbb
M}(x,\lambda, \mu) &=& T(0, x, \lambda) r_{ab}(\lambda -\mu) T(x,
-L, \lambda) K^-(\lambda) \hat T(0, -L, \lambda) \cr
 &+& T(0,
-L, \lambda) K^-(\lambda) \hat T(x, -L, \lambda) r_{ab}(\lambda
+\mu) \hat T(0, x, \lambda).
\label{mm}
\ea
Finally, bearing in mind the definition of $t(\lambda)$, and (\ref{last}) we conclude
\be
\Big \{ \ln\ t(\lambda),\ {\mathbb U}(x, \mu) \Big \} = {\partial {\mathbb
V}(x,\lambda, \mu) \over \partial x} + \Big [ {\mathbb V}(x,\lambda, \mu),\
{\mathbb U}(x, \mu) \Big ],
\label{defin}
\ee
where
\be {\mathbb V}(x,\lambda,
\mu) = t^{-1}(\lambda) \ tr_a \Big ( K_a^+(\lambda)\ {\mathbb M}_a(x , \lambda, \mu) \Big ).
\label{final1}
\ee
This is the explicit form for the $\mathbb{V}$-operator in the case of generic integrable boundary conditions.
One expands $\mathbb{V}$ in powers of $\l$ in order to obtain the modified operator
associated to each integral of motion of the model under consideration.

\subsection{Modified Lax pairs for the L-L model}
We are now in the position to determine the boundary Lax pair for boundary L-L model.
The classical $r$-matrix associated to the L-L model is proportional
to the permutation operator, and $tr_a \mathcal{P}_{ab} = \mathbb{I}$,
then $\mathbb{V}$ can be expressed in a simple form as
\ba
\mathbb{V}(x,\l,\m) & = &
\frac{t^{-1}(\l)}{\l-\m} T(x,-L,\l) K^-(\l) T^{-1}(0,-L,-\l) K^+(\l) T(0,x,\l)\cr
 & + & \frac{t^{-1}(\l)}{\l+\m} T^{-1}(0,x,-\l) K^+(\l) T(0,-L,\l) K^-(\l) T^{-1}(x,-L,-\l).
\label{LL_V1}
\ea
The latter expression is valid for all classical models associated to the Yangian classical $r$-matrix,
proportional to the permutation operator.

Explicit computation shows that the ${\mathbb V}$-operator for any point $x \neq 0,\ -L$ reduces to the familiar bulk operator (\ref{V_bulk}). In any case, we are mostly interested in computing the $\mathbb{V}$-operator exactly at the boundary points, that is $x_b = (0,-L)$. We
shall only present the final results here, and postpone the heavy technical details of the computation until the
Appendix C. At the end points the $\mathbb{V}$-operator has the following
form (we have multiplied the result of the expansion with ${1\over 2}$)
\be
{\mathbb V}_b(x_b) ={\mathbb V}(x_b) + \delta {\mathbb V}(x_b),
\ee
where ${\mathbb V}$ is the bulk operator (\ref{V_bulk}) and
\ba
\delta{\mathbb V}(-L) &=& {1\over 2 \mu} \left [ {\partial {\cal S} \over \partial x}  {\cal S} + {2\over i\xi^-} \begin{pmatrix}  {\mathrm k}^-(S^+(-L)- S^-(-L)) & -{\mathrm k}^- S_3(-L) -S^-(-L) \cr
{\mathrm k}^- S_3(-L) +S^+(-L)  & -{\mathrm k}^-(S^+(-L)- S^-(-L))
\end{pmatrix}\right ]
\cr
\delta{\mathbb V}(0) &=& {1\over 2 \mu} \left [ {\partial {\cal S} \over \partial x}  {\cal S} - {2\over i\xi^+} \begin{pmatrix}  {\mathrm k}^+(S^+(0)- S^-(0)) & -{\mathrm k}^+ S_3(0) -S^-(0) \cr
{\mathrm k}^+ S_3(0) +S^+(0)  & -{\mathrm  k}^+(S^+(0)- S^-(0)).
\end{pmatrix}\right ].
\ea
From the zero curvature condition, and by requiring $\delta{\mathbb V} =0$ (see more details on this argument in \cite{Avan:2007sf}), we obtain the equations of motion described in (\ref{e.o.m.}), and the 
non-trivial boundary conditions (see also \cite{Sklyanin:1987bi} for only diagonal boundary conditions):
\ba
&& \Big (S_2{\partial S_3 \over \partial x} - S_3{\partial S_2 \over \partial x}\Big )\Big \vert_{x=-L} =  {1 \over i\xi^-}S_2(-L) \cr
&& \Big (S_3{\partial S_1 \over \partial x} - S_1{\partial S_3 \over \partial x}\Big )\Big \vert_{x=-L} =-{1 \over i\xi^-}S_1(-L)- {2{\mathrm k}^- \over i\xi^-}S_3(-L) \cr
&& \Big (S_1{\partial S_2 \over \partial x} - S_2{\partial S_1 \over \partial x}\Big )\Big \vert_{x=-L} = {2 {\mathrm k}^- \over i\xi^-} S_2(-L).
\ea
Of course one may easily check that the same equations of motion, and boundary conditions are extracted from the Hamiltonian (\ref{bll}) via (\ref{emh}).
Note that in obtaining the boundary conditions we took into account that the derivative of the Casimir with respect to $x$ is zero. Note that similar equations of motion are obtained for the other end of the system ($x =0$), but are omitted here for brevity.
The entailed boundary conditions are as expected mixed ones.

\subsubsection{Dynamical boundary conditions}
The explicit computation of the modified ${\mathbb V}$-operator in this case 
follows exactly the previous section's computations via the expression 
(\ref{LL_V1}), so the result is quite straightforward, as long as we keep in 
mind that the classical dynamical reflection matrix is now expressed as in 
(\ref{dynamic}). Recall that we restrict our attention here to one boundary 
$x_b =-L$, then the final expression for the boundary operator is given as
\be
{\mathbb V}_b(-L) = {\mathbb V}(-L) + \delta{\mathbb V}(-L),
\ee
where we define
\be
\delta{\mathbb V}(-L)= {1\over 2\mu} \left [{\partial {\cal S} \over \partial x}  {\cal S} +
\begin{pmatrix}  {\mathbb Z}S^+(-L)- {\mathbb Y}S^-(-L) & -{\mathbb Z} S_3(-L) -2{\mathbb X}S^-(-L) \cr
{\mathbb Y}S_3(-L) +2{\mathbb X}S^+(-L)  & -{\mathbb Z}S^+(-L)+ {\mathbb Y}S^-(-L)
\end{pmatrix}\right ]
\ee
where ${\mathbb X},\ {\mathbb Y},\ {\mathbb Z}$ are defined in (\ref{opers}). In this case the relevant boundary conditions, entailed from the conditions $\delta{\mathbb V} =0$, read as
\ba
&& \Big (S_2{\partial S_3 \over \partial x} - S_3{\partial S_2 \over \partial x}\Big )\Big \vert_{x=-L} = {i ({\mathbb Z} -{\mathbb Y})\over 2}S_3(-L) +{\mathbb X} S_2(-L)\cr
&& \Big (S_3{\partial S_1 \over \partial x} - S_1{\partial S_3 \over \partial x}\Big )\Big \vert_{x=-L} =-{\mathbb X}S_1(-L)  -{{\mathbb Y} +{\mathbb Z} \over 2}S_3(-L) \cr
&& \Big (S_1{\partial S_2 \over \partial x} - S_2{\partial S_1 \over \partial x}\Big )\Big \vert_{x=-L} = { {\mathbb Y} +{\mathbb Z} \over 2} S_2(-L) -{i ({\mathbb Y} -{\mathbb Z}) \over 2} S_1(-L).
\ea
Needless to mention that these boundary conditions emerge also from the dynamical Hamiltonian (\ref{hdyn}) through (\ref{emh}).

\section{Integrable continuum limit}
We shall describe here a systematic means of obtaining classical continuum limits of quantum discrete theories for generic boundary conditions along the lines discussed in \cite{ADS}.
Assume a collection of operators assembled in matrices $L_{1i}$, acting on ``quantum''
Hilbert spaces labeled by $i$ and encapsulated
in a matrix ``acting'' on the auxiliary space $V_1$. For any quantum space $q$ they obey
the quadratic exchange algebra \cite{FTS, tak}
\be
R_{12}\ L_{1q}\ L_{2q} =L_{2q}\ L_{1q}\ R_{12} \ ,
\label{YBRgen}
\ee
where operators acting on different quantum spaces commute, and $R$ satisfies the Yang-Baxter equation.
The form of the monodromy matrix $T$ is then deduced from the co-module structure of the YB algebra
\be
T_a \equiv L_{aN}\ L_{a2}\ \ldots\ L_{a1}\ \label{TM1},
\ee
and thus naturally obeys the same quadratic exchange algebra (\ref{YBRgen}).

First consider that the $R$ matrix has a classical limit as
\be
R = 1 + \hbar r +{\cal O}(\hbar^2),
\ee
with $r$ satisfying the classical Yang-Baxter equation (see also e.g. \cite{sts,ADS}).
We may now establish that $T$ has a classical limit by considering in addition the classical counterpart of
$L$, which then satisfies the quadratic Poisson algebra
emerging directly as a semi-classical limit of (\ref{YBRgen}), after setting
$~{1\over \hbar} [A,\ B] \to \{ A,\ B \}$. It reads
\be
\{L_a(\lambda_1),\ L_b(\lambda_2) \} =
[r_{ab}(\lambda_1 -\lambda_2),\ L_a(\lambda_1)\ L_b(\lambda_2)]\ .
\label{semicl0}
\ee
The classical discrete monodromy matrix is apparently of the same form as in (\ref{TM1}).
The exchange algebra for $T$ takes the form
\be
\{T_a, T_b\} = [r_{ab},\ T_a\ T_b]\  .
\label{semicl}
\ee
This quadratic Poisson structure implies that the traces of powers of the monodromy
matrix $tr (T^c)$ generate Poisson-commuting quantities identified as classically integrable
Hamiltonians.

Now that we have discussed the classical limit we may proceed to the continuum 
limit of discrete theories with open boundary conditions. In this case the 
modified monodromy matrix has the form
\be
{\cal T}(\lambda) = T(\lambda)\ K^-(\lambda)\ T^{-1}(-\lambda), \label{tt2}
\ee
where $T$ is given by (\ref{TM1}), ${\cal T}$ satisfies the classical 
reflection equation (\ref{clrefl}) and $K^-$ is a $c$-number solution of the reflection equation (\ref{cnumber}).
Introduce a suitable spacing parameter $\delta:\ {\cal O}(\delta) \sim {\cal O}({1\over N})$.
Let us also express the $L$ matrix as
\ba
L_{an}(\lambda) &=& 1 + \delta {\mathbb U}_{an}(\lambda) + \delta^2 U^{(2)}_{an}(\lambda) + \ldots \cr
L^{-1}_{an}(-\lambda) &=& 1 - \delta {\mathbb U}_{an}(-\lambda) + \delta^2 \tilde U^{(2)}_{an}(-\lambda) + \ldots
\ea
It then naturally follows for the monodromy matrix and its inverse
\ba
T(\lambda) &=& 1 + \delta \sum_n  {\mathbb U}_{an}(\lambda) + \delta^2 \sum_{n>m}  {\mathbb U}_{an}(\lambda)  {\mathbb U}_{am}(\lambda) + \delta^2 \sum_n U_{an}^{(2)}(\lambda) + \ldots\cr
T^{-1}(-\lambda) &=& 1 - \delta \sum_n  {\mathbb U}_{an}(-\lambda)+ \delta^2 \sum_{n<m}  {\mathbb U}_{an}(-\lambda)  {\mathbb U}_{am}(-\lambda) + \delta^2 \sum_n \tilde U_{an}^{(2)}(-\lambda) + \ldots \nonumber\\
\label{mono}
\ea
For the following we consider that
\be
\delta \sum_j f_j \to \int_{-L}^0 dx\ f(x) ,~~~~~ {\mathbb U}_{aj} \to  {\mathbb U}_a(x), ~~~~ {\mathbb U}_{a j+1} \to  {\mathbb U}_a(x + \delta). \label{climit}
\ee
Based on the latter formulas it is clear that terms with powers of $\delta$ bigger 
than the number of summations go to zero in the continuum limit. This is the 
so-called ``power counting'' argument presented in more detail in \cite{ADS}. So the continuum limit of the monodromy matrix (\ref{mono}) becomes
\be
T(0, -L, \lambda) = {\mathrm P} \exp\{ \int_{-L}^0 dx\  {\mathbb U}(x) \}. \label{contt}
\ee
We conclude that the continuum limit of the discrete modified monodromy matrix reduces to the continuum analogue of (\ref{tt2}),  with $T$ given in (\ref{contt}).
The open transfer matrix, the generating function of the charges in involution as usual is
\be
t(\lambda) = tr_a \Big ( K^+(\lambda){\cal T}_a(\lambda) \Big ),
\ee
where $K^+$ a $c$-number solution of the classical reflection equation.
Having said these it is clear that continuum limits of discrete Hamiltonians would provide legitimate Hamiltonians of integrable continuum theories. In the two examples below we make this clear comparing also with the results of the the two previous sections.

\subsection{The open XXX chain}
It is quite straightforward now to check that the expression of the boundary Hamiltonian of the L-L model may be directly extracted
from the quantum XXX open chain as an appropriated continuum limit (see e.g. \cite{FR, ADS}). Recall the open XXX Hamiltonian \cite{Sklyanin:1987bi}, (see Appendix B for more details on the derivation):
\be
{\cal H} = \frac{1}{2} \sum_{j=1}^{N-1}\Big ( \sigma_j^x \sigma_{j+1}^x + \sigma_j^y \sigma_{j+1}^y
+ \sigma_j^z \sigma_{j+1}^z \Big ) - {i \over 2 \xi^-} \Big [ 2{\mathrm k}^-(\sigma_1^+ + \sigma_1^-) -\sigma_1^z\Big ]+ {i \over 2 \xi^+} \Big [ 2{\mathrm k}^+(\sigma_N^+ + \sigma_N^-) -\sigma_N^z\Big ].
\ee
The bulk term, after introducing appropriate coherent states see e.g. \cite{FR, ADS}, and after making the following identifications:
\be
\sigma^{x}_j \to S_{1}(x), ~~~~\sigma^y_j \to S_2(x), ~~~~\sigma^z_j \to S_3(x),~~~\mbox{and}~~~~\sigma_{j+1} \to S(x+\delta), \label{identif}
\ee
reduces to the bulk part of the Hamiltonian (\ref{bll}), whereas the boundary contributions of the left and right boundaries reduce exactly to the boundary terms of the Hamiltonian (\ref{bll}). Note that an implicit rescaling $\lambda \to \delta^{-1} \lambda$ takes place.
It is worth stressing here that continuum limits of discrete systems with boundary terms should be taken with particular care in order to have a sensible result. This is an intriguing issue also encountered in the case of integrable systems with local defects.

\subsubsection{Dynamical boundary conditions}
As in the non-dynamical case the classical Hamiltonian may be also directly derived as a continuum limit of the open XXX chain with a left dynamical boundary term (see also Appendix B):
\be
{\cal H} =\frac{1}{2} \sum_{j=1}^{N-1}\Big ( \sigma_j^x \sigma_{j+1}^x + \sigma_j^y \sigma_{j+1}^y
+ \sigma_j^z \sigma_{j+1}^z \Big ) +  \frac{1}{2}\sigma_1^+ \bar {\mathbb Z} +
\frac{1}{2} \sigma_1^- \bar {\mathbb Y}  -\frac{1}{2}\sigma_1^z \bar{\mathbb X} +\frac{1}{2}\bar {\mathbb D}, \label{hd2}
\ee
where we define
\ba
&& \bar {\mathbb X}= -4 ({\mathbb S}_3 +1) +{1\over i\xi^-} \Big (2 {\mathbb S}_3^2 +1  -2 {\mathrm k}^- \{{\mathbb S}^-,\ {\mathbb S}^3\} -2{\mathrm k}^- \{{\mathbb S}^+,\ {\mathbb S}_3\}\Big ) \cr
&& \bar {\mathbb Y} = 8{\mathbb S}^+ + {1\over i\xi^-} \Big ( 8{\mathrm k}^- ({\mathbb S}^+)^2 -2 {\mathrm k}^- ({\mathbb S}_3^2 -1) -2 \{{\mathbb S}_3,\ {\mathbb S}^+\} \Big ) \cr
&& \bar {\mathbb Z} = 8{\mathbb S}^- + {1\over i\xi^-} \Big ( 8{\mathrm k}^- ({\mathbb S}^-)^2 -2{\mathrm k}^-({\mathbb S}_3^2 -1) -2 \{{\mathbb S}_3,\ {\mathbb S}^-\} \Big )\cr
&& \bar {\mathbb D} = {4 \over i\xi^-} \Big (2{\mathrm k}^- ({\mathbb S}^+ + {\mathbb S}^-) - {\mathbb S}_3  \Big ), \label{dynel}
\ea
where $\{\ ,\ \}$ denotes the usual anti-commutator. Notice that the quantum dynamical ${\mathbb K}$-matrix is again of the form (\ref{kdyn}), however we considered for convenience the quantum Lax operator to be of the same structure as in (\ref{ll1}), but with an additional ${{\mathbb I} \over 2}$ term. The quadratic Casimir is chosen to be zero, so that the results are compatible with the classical continuum case. This means that we choose to consider the spin zero --non-compact-- representation of $\fr{sl}_2$.
Note also that we considered here a trivial right boundary $K^+ \propto {\mathbb I}$, a dynamical $K^+$ matrix for the right boundary would lead to extra boundary terms in expressions (\ref{hd2}) at $x=0$
of exactly the same from of as the ones at $x=-L$, but with $\xi^- \to -\xi^+,\ {\mathrm k}^- \to {\mathrm k}^+$ in (\ref{dynel}).

The boundary term of the quantum Hamiltonian above reduces to the boundary term of the classical continuum Hamiltonian, ensuring the consistency of the whole process with integrable continuum limits. There are some extra terms, which can be seen as ``quantum corrections'', and at the classical limit they vanish.

\section{Generalized boundary L-L models}
\subsection{The boundary anisotropic $\mathfrak{sl}_2$ L-L model}
This section serves mostly as a further check on the consistency of the continuum limit process we considered in the previous section. In particular, we shall start with the open XXZ Hamiltonian with generic boundary terms, and then we shall appropriately take the continuum limit in order to obtain the corresponding continuum Hamiltonian.

Let us first recall the associated $R$-matrix:
\be
R(\lambda) = \begin{pmatrix}
 \sinh(\l +{\mu \sigma^z \over 2} +{\mu \over 2}) & \sinh(\mu)\sigma^-
 \cr \sinh(\mu)\sigma^+ &\sinh(\l -{\mu \sigma^z \over 2}+{\mu \over 2})
\end{pmatrix},
\ee
The generic reflection matrices, associated to left and right boundaries, \cite{ghoza}, are given in Appendix B.

The Hamiltonian is then defined in Appendix B (see eq. (\ref{hqu})), and in a more explicit form is expressed as:
\ba
{\cal H}&=&\frac{1}{2} \sum_{j}\Big (\sigma_j^x \sigma_{j+1}^x + \sigma_{j}^y \sigma_{j+1}^y + \cosh(\mu) \sigma_{j}^z \sigma_{j+1}^z \Big ) + {\sinh (\mu) \over 2 \sinh (i\xi^- )} \Big (-\cosh(i\xi^-)\sigma_1^z + 2{\mathrm k}^- \sigma^x_1 \Big )\nonumber\\
& -&  {\sinh (\mu) \over 2 \sinh (i\xi^+ )} \Big (-\cosh(i\xi^+)\sigma_N^z + 2{\mathrm k}^+ \sigma^x_N \Big ).
\ea
Now recalling the identifications (\ref{identif}), the arguments presented in \cite{ADS}, and setting
\be
\sinh (\mu) = \delta \sqrt{J}+..., ~~~~~\cosh( \mu) = 1 -\delta^2 {J\over 2}+...
\ee
we may take the continuum limit of the latter expression and obtain the following classical Hamiltonian (see also \cite{Sklyanin:1987bi}):
\ba
{\cal I}_1 &=& -{1\over 4} \int_{-L}^0 dx \Big ( \left ({\partial \vec S \over \partial x}\right )^2 +J S_3^2  \Big) + {\sqrt{J}\over 2 \sinh(i\xi^-)} \Big (-\cosh(i\xi^-) S_3(-L) + 2 {\mathrm k}^- S_1(-L)\Big )\nonumber\\
&-&  {\sqrt{J}\over 2 \sinh(i\xi^+)} \Big (-\cosh(i\xi^+) S_3(0) + 2 {\mathrm k}^+ S_1(0)\Big ).
\ea
Utilizing the exchange relations among the classical spin variables we end up to the typical bulk equations of motion
\ba
&& \dot S_1 = i \Big ( S_2 {\partial^2 S_{3} \over \partial x^2}-S_3 {\partial^2 S_{2} \over \partial x^2}\Big )- i J S_2 S_3 \cr
&& \dot S_2 = i \Big ( S_3 {\partial^2 S_{1} \over \partial x^2}-S_1 {\partial^2 S_{3} \over \partial x^2} \Big )+ i J S_3 S_1 \cr
&& \dot S_3 = i \Big ( S_1 {\partial^2 S_{2} \over \partial x^2}-S_2 {\partial^2 S_{1} \over \partial x^2}\Big )
\ea
and the associated mixed boundary conditions (see also \cite{Sklyanin:1987bi}):
\ba
&& \Big (S_2{\partial S_3 \over \partial x} - S_3{\partial S_2 \over \partial x}\Big )\Big \vert_{x=-L} = {\cosh(i\xi^-)\sqrt{J} \over \sinh(i\xi^-)}S_2(-L) \cr
&& \Big (S_3{\partial S_1 \over \partial x} - S_1{\partial S_3 \over \partial x}\Big )\Big \vert_{x=-L} =-{\cosh(i\xi^-)\sqrt{J} \over \sinh(i\xi^-)}S_1(-L)-{2{\mathrm k}^- \over \sinh (i\xi^-)} S_3(-L) \cr
&& \Big (S_1{\partial S_2 \over \partial x} - S_2{\partial S_1 \over \partial x}\Big )\Big \vert_{x=-L} = {2{\mathrm k}^- \sqrt{J}\over \sinh (i\xi^-)} S_2(-L).
\ea
It is clear that in the isotropic limit we recover the equations of motion, and the associated boundary
conditions derived in  the previous section. Apparently analogous boundary conditions emerge for $x=0$, which are omitted here for brevity.

Having checked once more the consistency of the continuum limit process, we may now proceed with the $\mathfrak{gl}_n$ generalizations of the isotropic and anisotropic (trigonometric) L-L model.

\subsection{The boundary isotropic $\mathfrak{gl}_n$ L-L model}
We may now generalize our analysis for the isotropic $\mathfrak{gl}_n$ model. The easiest way to obtain the desired results is to start with the Hamiltonian of the corresponding quantum spin chain, and take the suitable continuum limit. This is an absolutely legitimate way of extracting continuum integrable Hamiltonians as has been transparent from the description presented in the preceding sections.

Let us first introduce the $R$ and $K$ matrices associated to the $\mathfrak{gl}_n$ model. The $\mathfrak{gl}_n$ $R$ matrix is given as
\be
R(\lambda) = \lambda +{\cal P} ~~~~~\mbox{where} ~~~~~{\cal P} =  \sum_{i,\ j=1}^n e_{ij} \otimes e_{ji}, \label{pp}
\ee
where we define: $(e_{ij})_{kl} = \delta_{ik} \delta_{jl}$. The $K^{\pm}$ matrices
in this case are of the generic form (see also \cite{annecy}):
\ba
&& K^-(\lambda,\ \xi^-,\ {\mathrm k}^-) = {\mathbb I} + \lambda \mathbb{B}^-, ~~~~~{\mathbb B}^- = {1\over i \xi^-}\Big ( -c_1^-e_{11} -c_2^- e_{nn} +2 {\mathrm k}^-(e_{1n} +e_{n1}) +c {\mathbb I}\Big ) \cr
&& K^+(\lambda,\ \xi^+,\ {\mathrm k}^+) = {\mathbb I} - (\lambda+1) \mathbb{B}^+, ~~~~~{\mathbb B}^+ = {1\over i \xi^+}\Big (
-c_1^+e_{11} -c_2^+ e_{nn} +2 {\mathrm k}^+(e_{1n} +e_{n1}) +c {\mathbb I}\Big )\cr
&& \mbox{where} ~~~~c_1^{\pm} = c^{\pm} + 1, ~~~~c_2^{\pm}=c^{\pm}-1 ~~~~c = 4 {\mathrm k}^{\pm 2} +1.
\ea
Then the boundary quantum discrete Hamiltonian is given as (see also Appendix B):
\ba
{\cal H} &=& \sum_{i,j=1}^n \sum_{m=1}^{N-1} e_{ij}^{(m)} e_{ji}^{(m+1)} +{1\over 2i\xi^-} \Big (-c_1^-e_{11}^{(1)} -c_2^- e_{nn}^{(1)} +2 {\mathrm k}^- (e_{1n}^{(1)} +e_{n1}^{(1)}) \Big ) \cr
&-&{1\over 2i\xi^+} \Big (-c_1^+e_{11}^{(N)} -c_2^+ e_{nn}^{(N)} +2 {\mathrm k}^+ (e_{1n}^{(N)} +e_{n1}^{(N)}) \Big ). \label{hh3}
\ea
Consider now the following identifications (see also \cite{ADS}),
\be
e_{ij}^{(m)} \to l_{ij}(x),  ~~~~e_{ij}^{(m+1)} \to l_{ij}(x+\delta) \label{ident2}
\ee
where the elements $l_{ij}$ satisfy the classical $\mathfrak{gl}_n$ algebra:
\be
\Big \{l_{ij}(x),\ l_{kl}(y) \Big \} =  \Big ( \delta_{il}l_{jk}(x) -\delta_{jk}l_{il}(x)\Big ) \delta(x-y).
\ee
We then  conclude about the continuum limit of the expression (\ref{hh3}):
\ba
{\cal I}_1 &=& -{1\over 2} \int_{-L}^0 dx\ \sum_{i,j =1}^nl'_{ij}(x) l'_{ji}(x) + {1\over 2i\xi^-} \Big ( -c_1^-l_{11}(-L) -c_2^-l_{nn}(-L) +2 {\mathrm k}^- (l_{1n}(-L) + l_{n1}(-L))\Big )\cr
&-&{1\over 2i\xi^+} \Big ( -c_1^+l_{11}(0) -c_2^+l_{nn}(0) +2 {\mathrm k}^+ (l_{1n}(0) + l_{n1}(0))\Big )
\ea
where the prime denotes derivative with respect to $x$.

From the latter expression the equations of motion and the corresponding generic mixed boundary conditions for the model are readily entailed:
\be
\dot{l}_{kl}(x) =  \sum_{j=1}^n \Big ( l''_{jl}(x) l_{jk}(x) -   l''_{kj}(x) l_{jl}(x) \Big )
\ee
and
\ba
&& \sum_{j=1}^n \Big ( l'_{jl}(-L) l_{jk}(-L) - l'_{kj}(-L) l_{jl}(-L) \Big )= -{1\over 2i\xi^-} \Big [ -c_1^-\Big (\delta_{1l} l_{1k}(-L) -\delta_{1k} l_{1l}(-L) \Big ) \cr && -c_2^- \Big (\delta_{nl} l_{nk}(-L)-\delta_{nk} l_{nl}(-L)\Big ) +2 {\mathrm k}^- \Big  ( \delta_{ln}l_{1k}(-L) - \delta_{1k} l_{nl}(-L) +\delta_{nl} l_{1k}(-L) - \delta_{ik} l_{nl}(-L)\Big ) \Big]. \nonumber\\
\ea
Analogous boundary conditions are clearly  obtained for $x=0$.

\subsection{The boundary anisotropic $\mathfrak{gl}_n$ L-L model}
We shall finally describe the anisotropic (trigonometric) L-L model starting from the ${\mathfrak U}_q(\mathfrak{gl}_n)$ open quantum spin chain. The $R$-matrix associated to the spin chain under consideration is given by \cite{jimbo}:
\be
R(\lambda) = {\cal P} \Big ( \sinh (\lambda +\mu)\ {\mathbb I} +\sinh (\lambda)\ { \mathrm U} \Big ),
\ee
where ${\cal P}$ is defined in (\ref{pp}), ${\mathrm U}$ provides a representation of the Hecke algebra (see also \cite{jimbo}), and is defined as
\be
{\mathrm U} = \sum_{i\neq j =1}^n \Big (e_{ij} \otimes e_{ji} - q^{-sgn(i-j)}e_{ii} \otimes e_{jj} \Big )
\ee
where $q = e^{\mu}$.
We consider the reflection matrix written in terms of the representation of the ``boundary'' element ${\mathrm e}$ of the 
affine Hecke algebra (see e.g.  \cite{doikou2})
\be
K^{\pm}(\lambda) =  x^{\pm}(\lambda)\ {\mathbb I} + y^{\pm}(\lambda)\ {\mathrm e}^{\pm},
\ee
where we define
\ba
&& x^-(\lambda)= -\delta_0^- \cosh(2\lambda +\mu)-\kappa^- \cosh(2\lambda)-2\cosh(2i\zeta^-)\sinh(\mu), \cr && y^-(\lambda) =
2 \sinh(2\lambda) \sinh (\mu) \cr
&& x^+(\lambda)= -\delta_0^+ \cosh(-2\lambda -\mu)-\kappa^+ \cosh(-2\lambda-2\mu)-2\cosh(2i\zeta^+)\sinh (\mu),
\cr && y^+(\lambda) =-2 \sinh(2\lambda+2\mu) \sinh (\mu) \cr
&& \delta_0^{\pm} = -(Q^{\pm} +(Q^{\pm})^{-1}), ~~~~~\kappa^{\pm} = q (Q^{\pm})^{-1} + q^{-1} Q^{\pm},
\ea
and $Q^{\pm},\ \zeta^{\pm}$ are free boundary parameters.
Also, choose a particular representation provided by
\be
{\mathrm e}^{\pm} = -(Q^{\pm})^{-1} e_{11} -Q^{\pm} e_{nn} +e_{1n} + e_{n1}.
\ee
The quantum spin chain Hamiltonian may be written then in terms of the affine Hecke elements as (see also \cite{doikou2}):
\be
{\cal H} =  \sum_{j=1}^{N-1} {\mathrm U}_{j\ j+1} +C^-\ {\mathrm e}^-_1 + C^{+}\ {\mathrm e}^+_N,
\ee
where $C^{\pm} =\mp {4\sinh (\mu) \over Q^{\pm} - (Q^{\pm})^{-1} -2 \cosh (2i\zeta^{\pm})}$.
Again considering the identifications (\ref{ident2}), and expressing ${\mathrm U}$ as \cite{ADS}
\be
{\mathrm U} = {\cal P} -{\mathbb I} + \sum_{i\neq j}^n \Big (1 -q^{-sgn(i-j)} \Big ) e_{ii} \otimes e_{jj}
\ee
and
\be
\mu =\delta \alpha, ~~~~~q^{-sgn(i-j)} \sim 1 - sgn(i-j)\delta \alpha +{\delta^2 \alpha^2 \over 2},
\ee
we obtain the following continuum boundary Hamiltonian, which may be seen as a deformation of the isotropic $\mathfrak{gl}_n$ model presented in the previous section:
\ba
{\cal I}_1 &=& -{1\over 2}\int_{-L}^0 dx \Big ( \sum_{i,\ j =1}^n  l'_{ij}(x) l'_{ji}(x)
+\alpha^2 \sum_{i\neq j=1}^n l_{ii}(x) l_{jj}(x) +2 \alpha \sum_{i<j=1}^n (l_{ii}(x) l'_{jj}(x) -l_{jj}(x) l'_{ii}(x)) \Big )
\cr &+& \tilde C^- \Big ( -(Q^{-})^{-1} l_{11}(-L) -Q^- l_{nn}(-L) +l_{1n}(-L) + l_{n1}(-L) \Big ) \cr
&+& \tilde C^+ \Big ( -(Q^{+})^{-1} l_{11}(0) -Q^+ l_{nn}(0) +l_{1n}(0) + l_{n1}(0) \Big ),
\ea
where $\tilde C^{\pm} = \mp {4 \alpha \over Q^{\pm} - (Q^{\pm})^{-1} -2 \cosh (2i\zeta^{\pm})}$. From the latter expression and the $\mathfrak{gl}_n$ exchange relations the associated equations of motion are entailed
\ba
\dot{l}_{kl}(x) &=& \sum_{j =1}^n\Big ( l''_{jl}(x) l_{jk}(x) - l''_{kj}(x) l_{jl}(x) \Big ) - \alpha^2 \Big ( \sum_{j\neq l} l_{jj}(x) l_{lk}(x) - \sum_{j\neq k} l_{jj}(x) l_{kl}(x)\Big ) \cr
&+ &2\alpha \sum_{j<l}\ l'_{jj}(x) l_{lk}(x) - 2\alpha \sum_{j<k} l'_{jj}(x)l_{kl}(x)
-2\alpha \sum_{j>l} l'_{jj}(x)l_{lk}(x) + 2\alpha \sum_{j>k} l'_{jj}(x)l_{kl}(x), \nonumber\\
\ea
with corresponding boundary conditions
\ba
&&\sum_{j =1}^n\Big ( l'_{jl}(-L) l_{jk}(-L) - l' _{kj}(-L) l_{jl}(-L) \Big ) =  -\alpha \sum_{j<l} l_{jj}(-L)l_{lk}(-L) \cr && + \alpha \sum_{j<k}l_{jj}(-L)l_{kl}(-L) +\alpha \sum_{j>l}l_{jj}(-L)l_{lk}(-L) - \alpha \sum_{j>k} l_{jj}(-L)l_{kl}(-L) \cr
&& -\tilde C^- \Big [  -Q^{-1}\Big(\delta_{1l}l_{1k}(-L) - \delta_{1k} l_{1l}(-L)\Big ) - Q\Big(\delta_{nl} l_{nk}(-L) - \delta_{nk} l_{nl}(-L)\Big )\Big ]\cr
&& -\tilde C^- \Big [ \delta_{1l} l_{nk}(-L) - \delta_{nk}l_{1l}(-L) + \delta_{nl} l_{1k}(-L) - \delta_{1k} l_{nl}(-L)\Big ],
\ea
and similarly for the other end of the theory at $x=0$.
With this we conclude our analysis on the generalized $\mathfrak{gl}_n$ boundary L-L models. Note that similar generalizations can be applied in a straightforward manner in the elliptic case, but are omitted here for brevity.

\section{Boundary symmetries}
We focus here mainly on the isotropic $\mathfrak{gl}_n$ case, and briefly discuss the continuum analogues of earlier woks on
boundary symmetries (see e.g \cite{doikou2}), although we have to mention that this is a whole separate subject of interest.

We shall extract below the so-called boundary non-local charges which are realizations of the underlying classical reflections algebra. We shall first consider $K^+ \propto {\mathbb I}$, and $K^-$ provided by the generic expression:
\be
K^-(\lambda) = {1 \over \lambda} + {\mathbb B}.
\ee
The $L$ and $L^{-1}$ operators are expressed as
\ba
L_{0i}(\lambda) &=& 1 +{\delta \over \lambda} {\mathbb P}_{0i} \cr
L^{-1}_{0i}(-\lambda) &=& 1 + {\delta \over \lambda} {\mathbb P}_{0i} + {\delta^2\over \lambda^2} {\mathbb P}_{0i}^2 + \ldots
\ea
We now consider the modified monodromy matrix ${\cal T}$, and by expanding in powers 
of ${1\over \lambda}$ we extract the boundary non-local charges (see also e.g. \cite{doikou2}). Let us start with the discrete ${\cal T}$
\be
{\cal T}_0(\lambda) = L_{0N}(\lambda) \ldots L_{01}(\lambda)\ K^-(\lambda)\ L^{-1}_{01}(-\lambda) \ldots L^{-1}_{0N}(-\lambda),
\ee
the expansion of the latter leads to
\ba
{\cal T}_0(\lambda) &=& {\cal T}_0^{(0)} +{1\over \lambda} {\cal T}_0^{(1)}+{1\over \lambda^2} {\cal T}_0^{(2)} + \ldots
\cr &=& {\mathbb B}_0 + {\delta \over \lambda} \Big ( \sum_i {\mathbb P}_{0i}{\mathbb B}_0 + {\mathbb B}_0 \sum_i {\mathbb P}_{0i}+1\Big ) \cr
&+&{\delta^2 \over \lambda^2}\Big (\sum_{i>j} {\mathbb P}_{0i}{\mathbb P}_{0j}{\mathbb B}_0  +{\mathbb B}_0\sum_{i<j} {\mathbb P}_{0i}{\mathbb P}_{0j} + {\mathbb B}_0 \sum_i {\mathbb P}_{0i}^2 + \sum_{i,j}{\mathbb P}_{0i}{\mathbb B}_0{\mathbb P}_{0j} + \sum_j {\mathbb P}_{0i}\Big ) + \ldots \nonumber\\ \label{charges}
\ea
To obtain the continuum limit of the latter expression we recall formulas (\ref{climit}) as well as the ``power counting'' argument, then
\ba
&& \delta \sum_i {\mathbb P}_{0i} \to \int_{-L}^0 dx\ {\mathbb P}_0(x), \cr
&& \delta^2 \sum_{i,j } {\mathbb P}_{0i} {\mathbb P}_{0j} \to \int dx\ dy\ {\mathbb P}_{0}(x){\mathbb P}_{0}(y). \label{contin2}
\ea
The modified monodromy matrix as well as the  ${\mathbb B},\ {\mathbb P}$ matrices may be expressed as
\ba
{\cal T} &=& \sum_{k,l=1}^n e_{kl} {\cal T}_{kl} \cr
{\mathbb P} &=& \sum_{k,l=1}^n e_{kl} {\mathbb P}_{kl} \cr
{\mathbb B} &=& \sum_{k,l=1}^n e_{kl} {\mathbb B}_{kl}. \label{matrixform}
\ea
The entries ${\cal T}_{kl}$ of the matrix are the non-local charges, which are realizations of the classical reflection algebra (\ref{clrefl}), (see also \cite{avandoikou2}). In particular, the continuum non-local charges emerging from (\ref{charges}) via (\ref{contin2}) become:
\ba
{\cal T}^{(0)}_{kl} &=& {\mathbb B}_{kl}, \cr
{\cal T}^{(1)}_{kl} &=& \sum_{m=1}^n \int_{-L}^0 dx\ {\mathbb P}_{km}(x) {\mathbb B}_{ml} +  \sum_{m=1}^n {\mathbb B}_{km}\int_{-L}^0 dx\ {\mathbb P}_{ml}(x),\cr
{\cal T}^{(2)}_{kl} &=& \sum_{m, p=1}^n \int_{x>y} dx\ dy\ {\mathbb P}_{km}(x) {\mathbb P}_{mp}(y){\mathbb B}_{pl} + \sum_{m, p=1}^n{\mathbb B}_{km} \int_{x<y} dx\ dy\ {\mathbb P}_{mp}(x) {\mathbb P}_{pl}(y) \cr &+& \sum_{m, p=1}^n\int_{x, y} dx\ dy\ {\mathbb P}_{km}(x) {\mathbb B}_{mp} {\mathbb P}_{pl}(y), \  \ldots \label{clexp}
\ea

Consider now the continuum expression for ${\cal T}$ with $T$ given in (\ref{contt}), and take into account that for the L-L model
\be
{\mathbb U}(x) = {1\over \lambda} {\mathbb P}(x),
\ee
where ${\mathbb P}(x)= \sum_{a,b=1}^n e_{ab} {\mathbb P}_{ab}(x)$, and the entries ${\mathbb P}_{ab}(x)$ satisfy the classical $\mathfrak{gl}_n$ algebra. It is then clear that expansion of the continuum ${\cal T}$ in powers of $1\over \lambda$ would exactly lead to expressions (\ref{clexp}) (see also the relevant discussion in section 5). Once more the consistency of the continuum limit is manifest.

We may now show that ${\cal T}_{kl}^{(1)}$, which form a closed algebra themselves, provide an exact symmetry of the continuum open transfer matrix (see also \cite{done, doikou2}), i.e.
\be
\Big [t(\lambda),\ {\cal T}^{(1)}_{kl} \Big ] =0.
\ee
We begin our proof by considering the classical reflection algebra (\ref{clrefl}) as $\lambda_1 \to \infty$:
\ba
\Big \{{\mathbb B}_a + {1\over \lambda_1} {\cal T}^{(1)}_a + \ldots,\ {\cal T}_b(\lambda_2) \Big \} &=& {1\over \lambda_1 -\lambda_2} \Big [{\cal P}_{ab},\ \Big ({\mathbb B}_a + {1\over \lambda_1} {\cal T}^{(1)}_a + \ldots \Big ){\cal T}_b(\lambda_2) \Big ] \cr
&+& {1\over \lambda_1 +\lambda_2} \Big ({\mathbb B}_a + {1\over \lambda_1} {\cal T}_a^{(1)} +\ldots \Big ) {\cal P}_{ab} {\cal T}_b(\lambda_2) \cr
&-& {1\over \lambda_1 +\lambda_2} {\cal T}_b(\lambda_2) {\cal P}_{ab} \Big ( {\mathbb B}_a +{1\over \lambda_1} {\cal T}_a^{(1)} +\ldots \Big ).
\ea
Keep only the first order terms ${1 \over \lambda_1}$ in the latter formula (set also $\lambda_2 = \lambda$), and bear in mind that
$[{\mathbb B}_a,\ {\cal T}_b] =0$ then,
\ba
&& \Big \{ {\cal T}^{(1)}_a,\ {\cal T}_b(\lambda) \Big \} = {\cal P}_{ab} {\mathbb B}_a{\cal T}_b(\lambda) - {\mathbb B}_a {\cal T}_b(\lambda) {\cal P}_{ab} + {\mathbb B}_a {\cal P}_{ab} {\cal T}_b(\lambda) - {\cal T}_b(\lambda) {\cal P}_{ab} {\mathbb B}_a\ \Rightarrow \cr
&&\Big \{ {\cal T}^{(1)}_a,\ tr_b \Big ({\cal T}_b(\lambda)\Big ) \Big \} = \ldots \ = 0\  \Rightarrow\ \Big \{ {\cal T}^{(1)}_{kl},\ t(\lambda) \Big \} = 0
\ea
and this concludes our proof on the exact symmetry of the transfer matrix.
Note that the indices $a,\ b$ refer to the auxiliary spaces, whereas the indices $k,\ l$ denote entries of the matrix according to
(\ref{matrixform}).

In fact, the above algebra, formed by ${\cal T}^{(1)}_{kl}$, is the full $\mathfrak{gl}_n$ algebra for $K^{-} \propto {\mathbb I}$ \cite{doikou2}, and it breaks down to suitable subalgebras depending on the choice of $K^{\pm}$. For instance, if we choose $K^{\pm}$ to be of the form:
\ba
&& K(\lambda) = diag \Big (\underbrace{a(\lambda), \ldots, a(\lambda)}_l,\ \underbrace{b(\lambda), \ldots ,b(\lambda)}_{n-l} \Big ), \cr
&& a(\lambda) =-\lambda+i\xi, ~~~~~b(\lambda) = \lambda + i\xi
\ea
then the exact symmetry reduces to $\mathfrak{gl}_l \otimes \mathfrak{gl}_{n-l}$ \cite{done, doikou2}.

Finally, a more generic choice of $K^+$ matrix would further reduce the exact symmetry of the transfer matrix, these issues have nevertheless been examined in e.g. \cite{done, doikou2}, and we shall not further discuss them here. Note also that the symmetry for anisotropic models is discussed in detail in \cite{doikou2}.

\section{Discussion}
Let us summarize the main findings of the present investigation: we have been able to 
explicitly derive expressions of the first classical integrals of motion for a 
prototype integrable model with non-trivial boundaries, that is 
the $\mathfrak{sl}_2$ isotropic Landau-Lifshitz model. We considered both $c$-number 
(non-dynamical), as well as dynamical classical reflection matrices, which give rise to 
non-trivial, but still integrable, boundary terms in the Hamiltonian of the 
model. Compatibility of our results with the classical limit of the corresponding quantum discrete model Hamiltonian, i.e. the XXX open spin chain, further ensures the validity of our findings. In addition to the integrals of motion we have been able to derive novel expressions for the associated boundary Lax pairs for the $\mathfrak{sl}_2$ isotropic L-L model, following the prescription introduced in \cite{Avan:2007sf}. As expected both the boundary Hamiltonians as well as the associated Lax pairs lead to the same equations of motion and boundary conditions, verifying the consistency of the methodology followed.

Having checked the consistency of the continuum limits of the XXX and XXZ open spin 
chains, leading to the isotropic and anisotropic $\mathfrak{sl}_2$ L-L model 
respectively, we then generalized our analysis to the isotropic and anisotropic 
$\mathfrak{gl}_n$ L-L models. More precisely, starting from the corresponding open 
spin chain Hamiltonians, with generic boundary terms, we considered the corresponding 
continuum limit, and extracted the associated classical boundary Hamiltonians 
as well as the equations of motion with the relevant boundary conditions. 
Finally, a brief discussion on the boundary symmetries is also presented. It is 
shown that the presence of special boundary terms suitably breaks the symmetry of 
the models under consideration, as also happens in discrete integrable 
models (see e.g. \cite{done, doikou2, avandoikou2}).

\paragraph{Acknowledgements\\}
NK acknowledges financial support provided by the Research Committee of the University of Patras
via a K. Karatheodori fellowship, under contract number C.915. He would also like to thank the
Physics Division of the National Technical University of Athens for kind hospitality during
the completion of this work.

\appendix

\section{Derivation of local integrals of motion}
We provide here the main technical points on the derivation of the local integrals of motion.
Recalling that the leading contribution of $e^Z$ and $e^{-\hat{Z}}$ as
$\l\to 0$ comes from the $Z_{11}$ and $\hat{Z}_{11}$ entries, we consider the following expansions
\ba
&&\Big[(1+\hat{W}(0,\l))^{-1}K^+(\l)(1+W(0,\l))\Big]_{11}=\sum_{n=0}^{\infty}\l^nh_n,\cr
&&\Big[(1+W(-L,\l))^{-1}K^-(\l)(1+\hat{W}(-L,\l))\Big]_{11}=\sum_{n=0}^{\infty}\l^n\bar{h}_n,\cr
&& \Big[Z(0,-L,\l)-\hat{Z}(0,-L,\l)\Big]_{11} =\frac{1}{\l}L
 + \sum_{n=1}^{\infty} (1-(-)^{n})\l^{n} \int_{-L}^0S^-(W_{n+1})_{21} dz.\cr
&&
\label{mod_exp}
\ea
Hence, the expansion of the modified transfer matrix is given by
\ba
&& \ln~tr\{K^+(\l)T(0,-L,\l)K^-(\l)\hat{T}(0,-L,\l)\} = \cr
&& \frac{1}{\l}L
 + \sum_{n=1}^{\infty} (1-(-)^{n})\l^{n} \int_{-L}^0S^-(W_{n+1})_{21} dz +
\ln\left(\sum_{n=0}^\infty\sum_{m=0}^{\infty}h_n\bar{h}_m\l^{n+m}\right).
\ea
However, we also need to expand the logarithm in the equation above. As will be clear below, the
first terms $h_0$ and $\bar{h}_0$ are proportional to unit. This fact enables us to write
the logarithm as
\be
\ln\left(\sum_{n=0}^\infty\sum_{m=0}^{\infty}h_n\bar{h}_m\l^{n+m}\right) = \ln\left(1 +
\sum_{n=1}^{\infty}(h_n+\bar{h}_n)\l^n + \sum_{n=1}^\infty\sum_{m=1}^{\infty}
h_n\bar{h}_m\l^{n+m}\right),
\ee
where the $h_n, \bar{h}_n$ that appear above have been rescaled as\footnote{This amounts to
adding the term $\ln(-\xi^+\xi^-)$ to the expansion of the modified transfer matrix.}
\be
h_n \to \frac{h_n}{h_0}, \qquad \bar{h}_n \to \frac{\bar{h}_n}{\bar{h}_0},
\label{h_rescal}
\ee
and expand thus the logarithm into powers of $\l$ as
\be
\ln\left(\sum_{n=0}^\infty\sum_{m=0}^{\infty}h_n\bar{h}_m\l^{n+m}\right) = \sum_{n=1}^\infty
f_n \l^n,
\ee
where $f_n$ provide essentially the boundary contributions to the integrals of motion for the
left and right boundary respectively. It is interesting to observe that the boundary
contribution decouples into terms associated with the left and right boundaries,
that is no mixing occurs
\ba
&& f_1 = h_1 + \bar{h}_1, \qquad f_2 = -\tfrac{1}{2}h_1^2 + h_2 -\tfrac{1}{2}\bar{h}_1^2 + \bar{h}_2\cr
&& f_3 = \tfrac{1}{3}h_1^3 - h_1h_2 +h_3 + \tfrac{1}{3}\bar{h}_1^3 - \bar{h}_1\bar{h}_2 +\bar{h}_3, \cr
&& \cdots
\ea
Next, we show explicitly how to compute the $h_n$'s,
that is the expansion of the first equation in \eqn{mod_exp}.

We now compute the inverse of $(1+\hat{W})$. Let
\be
(1+W)^{-1}=1+\sum_{n=0}^{\infty}\l^n F_n.
\ee
By demanding that the object above is  the inverse,
\be
(1+F_0 + \l F_1 + \l^2 F_2 + \cdots)(1 + W_0 + \l W_1 + \l^2 W_2 + \cdots) = 1,
\ee
we determine the quantities $F_n$:
\ba
F_0 & = & - W_0(1+W_0)^{-1} \cr
F_1 & = & -(1+F_0)W_1(1+W_0)^{-1}\cr
F_2 & = & -\left[(1+F_0)W_2 +F_1 W_1\right](1+W_0)^{-1}, ~~~~ \cdots
\ea
Recall also that $\hat{W}(\l)=W(-\l)$, then
\be
W_{2n+1}\to -W_{2n+1}, \qquad n=0,1,2,\cdots.
\ee
By denoting $(1 + \hat{W})^{-1}=1+\sum_{n=0}^{\infty}\l^n \hat{F}_n$, we find:
\ba
\hat{F}_0 & = & - W_0(1+W_0)^{-1} \cr
\hat{F}_1 & = & + (1 + \hat{F}_0)W_1(1+W_0)^{-1}\cr
\hat{F}_2 & = & -\left[(1 + \hat{F}_0)W_2 - \hat{F}_1 W_1\right](1+W_0)^{-1}.
\ea
Substituting the explicit forms of the relevant quantities we arrive at
\ba
&& h_0 = i \xi^+,\cr
&& h_1 = 2{\mathrm k}^+~S_1(0) - S_3(0)+i\xi^+ \left(\frac{4S^{+}(0)S^{-'}(0)}{1+S_3(0)}+S_3'(0)\right)\cr
&& =  2{\mathrm k}^+~S_1(0)-S_3(0)+\frac{\xi^+ }{1+S_3(0)}\Big (iS_3'(0) + S_1(0)S_2'(0) - S_1'(0)S_2(0) \Big ),
\ea
and
\ba
&& \bar{h}_0 = i\xi^-,\cr
&& \bar{h}_1 = -2{\mathrm k}^-~S_1(-L) + S_3(-L)-i\xi^-\left(\frac{4S^{+}(-L)S^{-'}(-L)}{1+S_3(-L)}
+S_3'(-L)\right)\cr
&& = -2{\mathrm k}^-~S_1(-L) + S_3(-L)-\frac{\xi^- }{1+S_3(-L)} \Big (iS_3'(-L) +
S_1(-L)S_2'(-L) - S_1'(-L)S_2(-L) \Big ) \cr
&&
\ea
These functions are not the rescaled  ones defined in \eqn{h_rescal}.
It is clear from the expansions of the left and right boundary contributions that the following symmetry
holds
\be
\bar{h}_n=(-1)^nh_n:~~~0\to-L,~~\xi^+\to \xi^-,~~{\mathrm k}^+ \to {\mathrm k}^-.
\ee

\section{The open spin chain Hamiltonian}
The open Hamiltonian emerges as a first derivative of the associated quantum transfer matrix of the model (see also e.g. \cite{Sklyanin:1987bi}). Note that the quantum transfer matrix for the boundary discrete model is of the same structure as the classical one (\ref{opentransfer}). Let us first consider the isotropic case
The final expression of the open quantum Hamiltonian is given by:
\be
{\cal H} \propto \sum_{j=1}^{N-1} H_{j\ j+1} + {1\over 2}{d {\mathbb K}^-_1(\lambda)\over d \lambda} \Big \vert_{\lambda=0} + {tr_0(K^+_0(0)H_{N0}) \over tr_0 K^+(0)} \label{hcl}
\ee
where we define for the isotropic ($\mathfrak{gl}_n$) case:
\be
H_{jk} = {\cal P}_{jk}.
\ee
Assume that the ${\mathbb K}^-$-matrix, solution of the quantum reflection equation, has the following generic form
\be
{\mathbb K}^- \propto {\mathbb I} + \lambda {\mathbb B} + {\cal O}(\lambda^2),
\ee
it is then clear that the boundary contribution is essentially the ${\mathbb B}$-matrix.

In the XXX case we have:
\ba
{\mathbb B}=
{1\over i \xi^-}\begin{pmatrix}
 -1 & 2{\mathrm k}^- \cr
 2{\mathrm k}^- & 1
\end{pmatrix} ~~~~~&&\mbox{$c$-number representation}\cr
{\mathbb B}=
\begin{pmatrix}
 -\bar {\mathbb X} +\bar {\mathbb D} & \bar {\mathbb Z} \cr
 \bar {\mathbb Y} & \bar {\mathbb X} +\bar {\mathbb D}
\end{pmatrix} ~~~~~&&\mbox{dynamical representation},
\ea
where the elements $\bar {\mathbb X},\ \bar {\mathbb Y},\ \bar {\mathbb Z},\ \bar {\mathbb D}$ for the dynamical case are defined in (\ref{dynel}). Also the right boundary is defined as:
\be
K^+(\lambda) \propto \begin{pmatrix}
 \lambda + 1+ i\xi^+ & -2{\mathrm k}^+(\lambda+1) \cr
 -2{\mathrm k}^+(\lambda+1) & -\lambda - 1 +i\xi
\end{pmatrix}
\ee

For the anisotropic case the generic expression one gets for the Hamiltonian is:
\be
H \propto \sum_{j=1}^{N-1} H_{j j+1} + {\sinh \mu \over 2 \sinh (i \xi^-)} {d K^-_1(\lambda)\over \d \lambda}\Big \vert_{\lambda =0}
+  {tr_0(K^+_0(0)H_{N0}) \over tr_0 K^+(0)} \label{hqu}
\ee
where we define
\be
H_{jj+1} = {\cal P}_{j j+1} {d R_{j j+1}(\lambda) \over d \lambda}\Big \vert_{\lambda =0}.
\ee
In the XXZ case in particular we choose:
\ba
K^-(\lambda) &=& \begin{pmatrix}
 \sinh (-\lambda+i\xi^-) & {\mathrm k}^-\sinh(2\lambda) \cr
 {\mathrm k}^-\sinh(2\lambda) & \sinh(\lambda +i\xi^-)
\end{pmatrix},
\nonumber\\
K^+(\lambda) &=& \begin{pmatrix}
 \sinh (\lambda +\mu +i\xi^+) & -{\mathrm k}^+\sinh(2\lambda+2\mu) \cr
 -{\mathrm k}^+\sinh(2\lambda+2\mu) & \sinh(-\lambda -\mu + i\xi^+)
\end{pmatrix}.
\ea

\section{The modified $\mathbb{V}$-operator}
To compute the modified $\mathbb{V}$-operator, we begin with \eqn{LL_V1}.
The monodromy matrix satisfies $T(x,x,\l) = T^{-1}(x,x,\l) = \mathbb{I}$, so  the results are greatly simplified
at the boundary points.

Substituting the ansatz \eqn{transatz} into
\eqn{LL_V1} and taking into account the explicit form of the transfer matrix presented
in Appendix A, \eqn{mod_exp}, one derives the exact expressions for the
$\mathbb{V}$-operator.

For the boundary point $x_b=0$ we have:
\be
\mathbb{V}_{ij} = X_R \left\{\frac{1}{\l-\m} A_{i1} B_{1j} + \frac{1}{\l+\m} C_{i1} D_{1j} \right\},
\ee
where we have defined
\ba
A & = & 1+ W(0),\qquad \qquad B = (1+\hat{W}(0))^{-1}K^+(\l), \cr
D & = & (1+ \hat{W}(0))^{-1},\qquad~~~  C = K^+(\l)(1+W(0)),
\ea
and $X_R$ is the quantity appearing in the first line of \eqn{mod_exp}. It is straightforward to
compute the expression for $\mathbb{V}$, since each one of the quantities above is
known.

In a similar fashion, for the boundary point $x_b = -L$:
\be
\mathbb{V}_{ij} = X_L \left\{\frac{1}{\l-\m} A_{i1} B_{1j} + \frac{1}{\l+\m} C_{i1} D_{1j} \right\},
\ee
where we have now defined
\ba
A & = & K^-(\l)(1 + \hat{W}(-L)),\qquad \qquad B = (1+ W(-L))^{-1}, \cr
D & = & (1 + W(-L))^{-1}K^-(\l),\qquad\qquad C = (1+\hat{W}(-L)),
\ea
and $X_L$ is now the quantity appearing in the second line of \eqn{mod_exp}.

\end{document}